**The Promise of Cross-Species Coexpression Analysis in Studying the Coevolution and Ecology of Host-Symbiont Interactions.**

Amanda K. Hund[1], Peter Tiffin[2], Jean-Gabriel Young[3], Daniel I. Bolnick[4]

1. University of Minnesota, Department of Ecology, Evolution, and Behavior, St. Paul, Minnesota, USA

2. University of Minnesota, Department of Plant and Microbial Biology, St. Paul, Minnesota, USA

3. University of Vermont, Department of Computer Science, Burlington, Vermont, USA

4. University of Connecticut, Department of Ecology and Evolutionary Biology, Storrs, Connecticut, USA

Corresponding Author: Amanda Hund, ahund@umn.edu

**Abstract**

Measuring gene expression simultaneously in both hosts and symbionts offers a powerful approach to explore the biology underlying species interactions. Such dual or simultaneous RNAseq approaches have primarily been used to gain insight into gene function in model systems, but there is opportunity to expand and apply these tools in new ways to understand ecological and evolutionary questions. By incorporating genetic diversity in both hosts and symbionts and studying how gene expression is correlated between partner species, we can gain new insight into host-symbiont coevolution and the ecology of species interactions. In this perspective, we explore how these relatively new tools could be applied to study such questions. We review the mechanisms that could be generating patterns of cross-species gene coexpression, including indirect genetic effects and selective filters, how these tools could be applied across different biological and temporal scales, and outline other methodological considerations and experiment possibilities.

**Introduction**

Most, if not all, species are involved in intimate and durable symbiotic interactions with individuals of other species (e.g., parasitism, host-mutualism). To understand the biology of such interactions, it is valuable to study the species together. Measuring gene expression simultaneously in both hosts and symbionts provides a powerful approach (Le Luyer et al. 2021; Du et al. 2022; Maulding et al. 2022). Multi-species gene expression (i.e. dual RNAseq or simultaneous transcriptomics) has primarily been used by molecular biologists and immunologists to address proximate questions about gene function and metabolic interactions between pathogens and the immune system (reviewed in (Westermann et al. 2012, 2017; Schulze et al. 2016; Marsh et al. 2017; Naidoo et al. 2018; Wolf et al. 2018; Baddal 2019; O'Keeffe and Jones 2019; Chung et al. 2021)). Moreover, studies using dual RNAseq approaches have typically focused on a single host and a single symbiont genotype, often of laboratory model systems, thus overlooking the role of genetic diversity within each interacting species. These studies also typically analyze gene expression (differential expression) in each interacting species separately, missing the opportunity to further evaluate interactions between species. We see substantial opportunity to apply and expand these studies by analyzing correlated gene expression across species jointly. The use of such cross-species gene expression and coexpression network analysis could help us better understand indirect genetic effects, selective filters that determine host and symbiont fitness, the functional basis of ecological interactions, and host-symbiont coevolution.

Genome-wide expression data identify the portion of the genome that is actively transcribed in specific environmental conditions in a given tissue and ontogenetic stage (Lovén et al. 2012). These data can be used to gain insight into the genes involved in development or real-time response to environmental conditions (Costa-Silva et al. 2017). Differential expression analyses typically compare the expression of each gene between different treatments, which could be different tissues, environments, genotypes, timepoints, or species, and are designed to identify which genes are differentially expressed due to these factors, or interactions between them. However, motivated by the understanding that genes do not act in isolation and that coexpressed genes are likely functionally related, researchers developed coexpression networks as an efficient means to synthesize and integrate information from many genes (van Dam et al. 2018; Kakati et al. 2019; Rao and Dixon 2019). Rather than examining the expression of each gene individually, coexpression networks highlight which pairs of genes are coexpressed. By 'coexpressed' we mean two genes' expression level (mRNA abundance) varies among biological samples (e.g., individuals), and their respective expression levels are correlated with each other. Groups of multiple genes may be coexpressed, which can be represented as modules within a coexpression network (Box 1). Coexpression analysis can capture variation along developmental trajectories or among environments, phenotypes, or genotypes. Studies describing coexpression modules have been key to identifying gene functions and regulatory pathways and their context-dependence (Amar et al. 2013; van Dam et al. 2018; Mishra et al. 2021).

Single-species coexpression networks have recently provided new insights into evolutionary patterns and processes. For example, comparative coexpression analysis has been used to characterize the extent to which phenotypic innovations and diversification have been accompanied not only by changes in gene expression (i.e. differential expression) but also by changes in transcriptome coexpression (i.e. network structure) (Swanson-Wagner et al. 2012; Filteau et al. 2013; Hu et al. 2016). Comparative coexpression analyses also have been used to identify key elements of gene expression that change or are conserved during species divergence (Pavey et al. 2010; Fruciano et al. 2019; Kolora et al. 2021; Ospina et al. 2021), as well as how gene expression changes enable populations to persist in novel environments, which can contribute to local adaptation and ecological speciation (Jacobs et al. 2020; Jacobs and Elmer 2021; Rajkov et al. 2021). Coexpression networks have provided insight into how interactions between genes shape evolutionary constraints and selection (Mack et al. 2019; Müller et al. 2019; Brown and Kelly 2022). These studies have found consistent support for more highly-connected genes experiencing greater selective constraint, although these genes may also contribute to adaptation in response to strong selection (Hämälä et al. 2020a). Finally, under the expectation that coexpression modules underlie variation in quantitative traits (Brown and Kelly 2022), coexpression networks can be used to test for evidence of adaptation on polygenic traits – signals of which might be too weak to be detected when genes are examined individually.

Here, we propose that cross-species gene expression data and coexpression network analyses can yield new insights into the mechanisms and coevolution of host-symbiont interactions. Studying partner species concurrently, rather than studying each species alone, allows researchers to assess genotype-specific physiological or metabolic activity, genotype by genotype variation, and will allow us to better quantify indirect genetic effects and cross-species epistasis. In this Perspective, we evaluate how cross-species gene expression and network analysis can be used to gain new insights into host-parasite coevolution and how ecology shapes these interactions. We propose that these approaches offer a powerful framework and toolkit for studying enduring ecological and evolutionary questions. For clarity, we focus much of our discussion around host-parasite interactions. However, the ideas presented apply to any system in which individuals of one species participate in sustained direct interactions with individuals of another species, such as host-mutualist interactions.

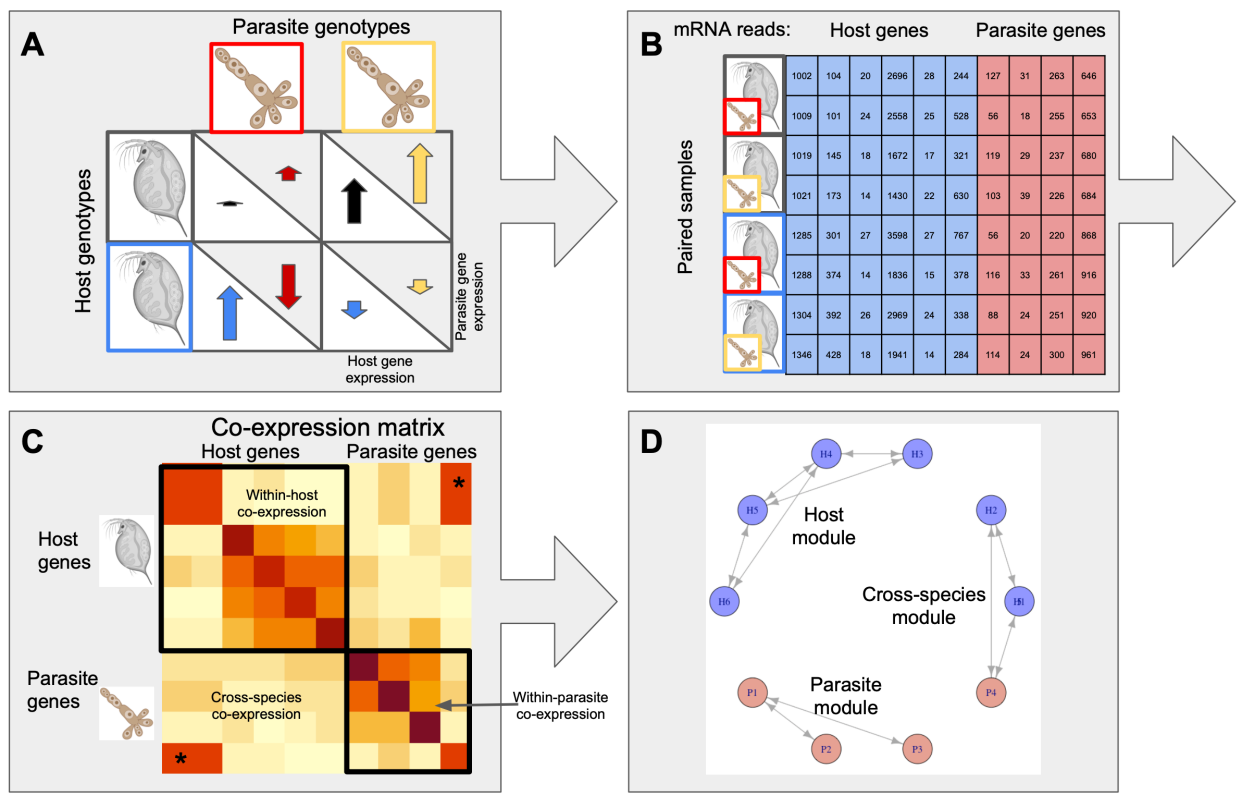

**Figure 1.** A schematic illustration of the idea of cross-species gene coexpression. A) Consider a genetically variable population of hosts (e.g., *Daphnia*), and of parasites (e.g., Metchnikowia yeast), in which all factorial combinations of host and parasite genotypes interact. Host gene expression (lower left triangles within each cell of the table) and parasite expression (upper right triangles) both vary as a function of host genotype, parasite genotype, and perhaps their interaction. **B)** If one were to sequence the transcriptomes of multiple pairs of *Daphnia* individuals and their parasites, the result would be a data table where host-parasite pairs are rows, host genes and parasite genes are separate columns, and one has read counts for each gene in each host-parasite pair. **C)** This data table can be converted into a correlation matrix with N rows and N columns, where N is the sum of the number of parasite genes $N_P$ and number of host genes $N_H$. The upper left $N_H*N_H$ square represents the set of correlations among host genes, with positive correlations in darker red. The lower right $N_L*N_L$ square represents the correlations among parasite genes. Correlations between host genes and parasite genes are found in the $N_H*N_L$ rectangle (a significant cross-species correlation is marked with an asterisk). **D)** By applying a threshold, this correlation matrix can be converted into an adjacency matrix of 0s (uncorrelated) and 1's (correlated) that may be represented as a network. This adjacency matrix can be subjected to standard network statistics analyses, for example, to detect modules of coexpressed genes (e.g., WGCNA software). These modules may be predominantly among genes within hosts, or within parasites, or they may include both species' genes representing cross-species coexpression modules. Figure created with BioRender.com and R.

**Cross-species Coexpression**

In this perspective, we define cross-species coexpression (CSCoE) as the correlated up- or down-regulation of genes expressed by interacting individuals of two (or more) species. CSCoE analyses require gene expression data from each of the interacting species (paired individuals) – in other words, for a host-parasite system, one needs to collect data on both among-host variation in parasite gene expression and among-parasite variation in host gene expression. These data may come from natural populations where both hosts and parasites are genetically polymorphic, generating various combinations of host and parasite genotypes, or, from experimental infections where multiple host genotypes are factorially infected with various parasite genotypes (Fig. 1A). These data provide a rich opportunity for expression, differential expression, and coexpression analyses – all of which can provide insight into the ecology and evolution of symbiotic interactions. The idea is illustrated in Figure 1, in which two host genotypes interact with two parasite genotypes. The level of expression in either species may depend on the genotype of the focal species, the genotype of the individual it interacts with, and perhaps their epistatic interaction (Fig. 1A). As a first pass analysis, each species' gene expression can be examined separately. Expression of a single gene *i* from parasite genotype *P* (represented here as $E_{i,P}$; the upper triangles in Fig. 1A) can be analyzed with linear models to test for the effects of parasite genotype ($G_P$), host genotype ($G_H$), and a host-parasite genotype interactions: $E_{i,P} \sim G_P + G_H + G_P \times G_H$. Genotype here could refer to variation at a particular locus or could represent different populations. A large effect of the parasite genotype implies heritable variation in expression within the parasite population, while a main effect of host genotype implies an indirect genetic effect on parasite gene expression (see next section). The interaction between host and parasite genotype represents an example of what one might call *cross-species epistasis*. While we focus on genetic effects, environmental variables, of course, could also be included in the linear model.

Characterizing gene expression and differential expression separately for each species in the interaction has been widely adopted in dual RNAseq studies (Westermann et al. 2012, 2017;

Naidoo et al. 2018; Wolf et al. 2018; Chung et al. 2021), yet the vast majority of studies stop there. To get at CSCoE additional steps are needed. Rather than test for an effect of host genotype on parasite expression, one can test whether the focal parasite gene expression depends on the expression level of a particular host gene *j* ($E_{j,H}$). That is, we can test for a model in which $E_{i,P} \sim G_P + E_{j,H} + G_P \times E_{j,H}$, where CSCoE is revealed by a main effect of host expression on parasite expression. These analyses can be done reciprocally for host expression as a function of parasite expression. Statistical software exists (e.g., DeSeq2,(Love et al. 2014)) to fit statistical linear models for individual genes, though more detailed experimental designs would be needed to establish the direction of causation in these relationships.

This linear model framework, however, is focused on a single gene in just one of the species. An appealing alternative is to treat the host and parasite as equal partners and switch to a global approach where we focus on the relationship between their entire transcriptomes. Given $N_H$ host genes and $N_P$ parasite genes, measured for M host-parasite pairs, one has a M x ($N_H$ + $N_P$) data matrix of gene expression (Fig. 1B). Each row is a host-parasite pair, with columns representing both host and parasite gene expression levels. One way we may capture CSCoE is by converting the expression data into a ($N_H$ + $N_P$) x ($N_H$ + $N_P$) dimensional gene-gene correlation matrix (Fig. 1C). One diagonal square within the matrix represents gene coexpression within the host, a second diagonal square represents gene coexpression within the parasite. Of particular interest to us here is the off-diagonal rectangle representing the $N_H$ by $N_P$ correlations between the $N_H$ hosts and $N_P$ parasite genes (Fig. 1C). These cross-species associations should generally be weak (see Box 1), except where expression of a particular gene in one species leads to physiological changes that modify the expression of the other species, for example when helminth parasites suppress host immune function. The coexpression matrix can be analyzed directly to look for gene-gene pairs with particularly strong coexpression between species.

As an alternative to inspecting correlation matrices directly, we may use tools of network analysis to further understand coexpression data (see Box 1). In such analyses, correlation matrices are interpreted as the adjacency matrix of a network, with coefficients indicating which pair of genes interact and (for weighted networks) how strongly they interact. This network structure can be subjected to a variety of analyses—such as understanding the overall robustness of the network to removal of genes, identifying cohesive modules, or isolating critical genes (Newman 2018). Because many of these analyses require an input in which interactions either exist or don't, correlations can then—but need not—be converted to a binary format based on p-values, backbone-filtration methods (Serrano et al. 2009), or more recent methods that can account for randomness in the sequencing process (Poisot et al. 2016; Young et al. 2021). Regardless of the method, the end result is a CSCoE network, weighted or binary, (Fig. 1D) which can be analyzed to identify key interactions (see Box 2). Different from classical gene interaction networks, the nodes of this network will belong to different species. The approach proposed here is an improvement over the bipartite networks often used to analyze host-parasite ecological interactions (Guimarães 2020), in which one describes connections between hosts and parasites, but does not consider host-host or parasite-parasite covariance.

Taking network approaches a step further, multilayer network analysis (Kivelä et al. 2014; Aleta and Moreno 2019; Hammoud and Kramer 2020) provides methods to represent and analyze such data, in which data from each species is represented as a separate network (layer), with inter-layer connections identifying correlations between the expression of genes from different species. These form a layer-disjoint network, in that each layer is a network with unique nodes (genes) for that layer's species. CSCoE may then be studied by identifying modules that include genes from different species (Fig. 1D), using multilayer visualization software such as MuxVIS (De Domenico et al. 2015), module detection methods (Peixoto 2015; De Bacco et al. 2017; Wilson et al. 2017; Huang et al. 2021) or inferential tools such as multilayer exponential random graph models (Chen 2021). As noted later in this paper, multi-layer network analyses allow one to incorporate more layers such as additional species, or other phenotypic data layers (e.g., proteomics). To our knowledge, few studies have applied network approaches to cross-species gene expression data, and tools for multilayer network analyses are a very active area of current progress.

The architecture of cross-species coexpression network(s) could provide information about the interdependencies of hosts and parasites. At one extreme, each of the network modules would contain only host or only parasite genes (and no 'inter-layer edges' in multilayer network lingo). Such a network structure would indicate that much of the expression variation within hosts or parasites is not directly shaped by gene expression in the partner species. The lack of inter-species edges may reflect the lack of genetic variation in host response to parasite infection (or parasite response to host genotype), for instance, if high host tolerance means that parasites have little effect on host physiology. At the other extreme, many of the network modules would contain genes from both species. This would suggest that the among-sample variation in gene expression in one species is closely linked to the expression of specific genes in the other species. Such cross-species modules would be particularly interesting if a gene from one partner was strongly connected to multiple genes of the other species. Such hub genes may be "keystone genes", paralleling the concept of keystone species which has an exceptionally large effect on the structure of its ecological community. Such genes could underlie species interactions and be the target of coevolution, for example, key effector proteins that enable infection or trigger a host response. Integration of additional information with gene coexpression networks, such as known regulatory pathways or transcription factor binding sites, could help with the construction of gene regulatory networks to better predict and understand molecular inter-species interactions (Schulze et al. 2016).

**Underlying mechanisms generating patterns of Cross-species coexpression**

There are a number of mechanisms, both evolutionary and ecological, that could generate observable patterns of CSCoE. Understanding these underlying mechanisms will be important when analyzing data and interpreting findings from CSCoE studies. Though we caution that teasing them apart will be difficult without carefully designed experiments and in-depth knowledge of the study system.

Coexpression between species could arise from indirect genetic effects (IGEs). IGEs occur when different genotypes of one species induce different phenotypic changes in an individual of another species (Anacleto et al. 2019; Baud et al. 2022; De Lisle et al. 2022). For example, (Guo et al. 2017), infected isogenic *Medicago truncatula* plants with genetically distinct lines of parasitic nematodes and found that parasite genotype had a substantial effect on plant gene expression, influencing the expression of >200 host genes, with individual parasite loci causing up to a 90-fold change in host gene expression levels. (Mateus et al. 2019) found in cassava and mycorrhizal fungi that a large percentage of both plant and fungal transcriptional responses changed in direction and magnitude depending on their partner's genotype. To further illustrate this idea, consider a hypothetical example in which a parasite harbors genetic variation in the ability to suppress host immunity (Fig 2) – the genetic variation in this suppression trait can lead to variation in both parasite and host phenotypes. The among-parasite phenotypic variation is a direct consequence of polymorphism within the parasite population whereas there can be variation among individual hosts of identical genotypes as an indirect consequence of heritable among-parasite variation in immune suppression, i.e. an IGE of parasites on hosts (Fig 2, columns). The reciprocal direction of cause and effect is also possible. Hosts may be polymorphic in their capacity to mount an effective immune response, which will indirectly impact parasite phenotype (Fig 2, rows). These causal directions can act simultaneously (reciprocal IGEs). In this case, the interaction between host and parasite genotypes can result in cross-species epistasis (GxG) in which the fitness (and expression) effect of one genotype depends on the genotype of the other species.

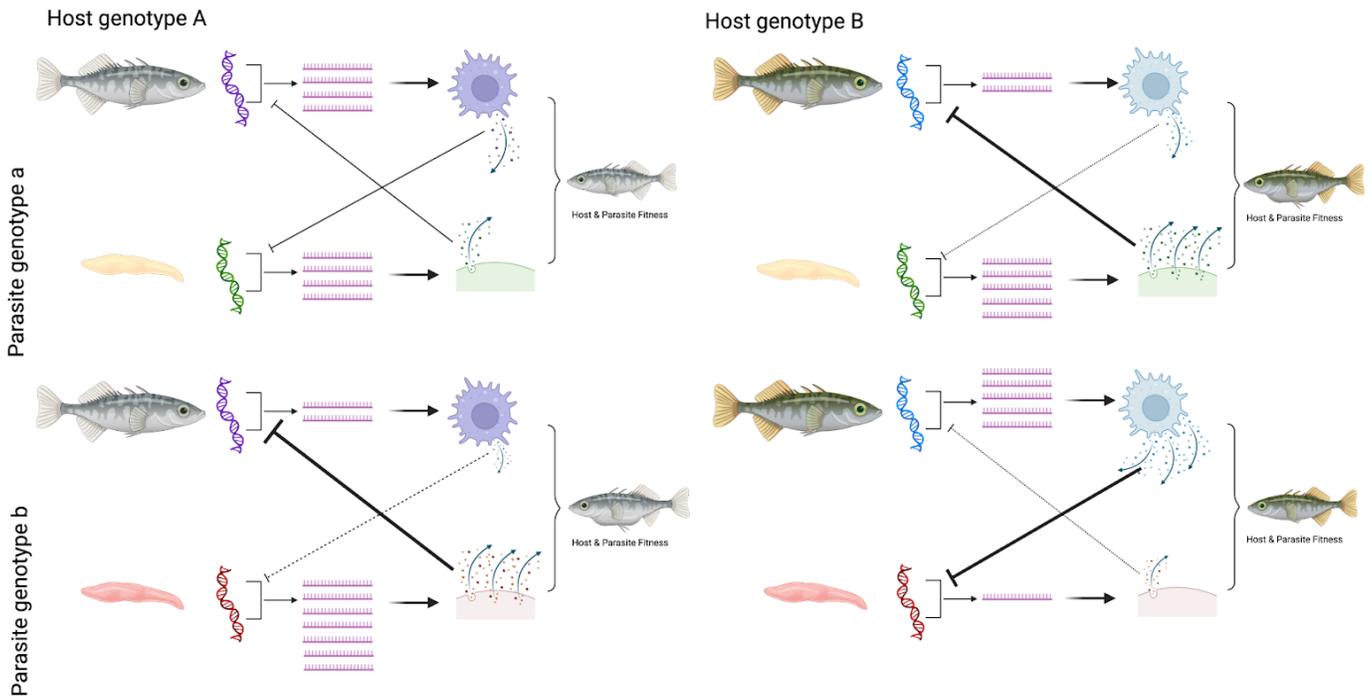

**Figure 2.** An illustration of how indirect genetic effects generate cross-species coexpression, and an experimental design strategy to test for IGEs. Consider two host genotypes (A, B), denoted by differently-shaded stickleback here, and two parasite genotypes (a,b) indicated by differently-colored cestodes. These may be defined by having different alleles at a particular locus or coming from different populations. For the particular combination of host A and parasite a (top left), the host genome encodes expression of mRNA of a particular gene that causes an immune cell to produce a product. The parasite genome produces mRNA of a particular gene that leads to a secreted compound that inhibits host gene expression

(e.g. parasite-mediated immune suppression). The host immune product in turn inhibits the parasite's ability to produce its secreted compound. Now, we will consider the consequence of varying either the host genotype while keeping the parasite genotype constant (rows) or varying the parasite genotype while keeping the host genotype constant (columns). When the parasite genotype expresses its immune suppression mRNA at a higher level within a certain host genotype, it can more effectively suppress host immune expression, resulting in a successful infection (lower left, top right). Conversely, in cases where host genotype expresses immune mRNA at a higher level given a certain parasite genotype, more immune product is generated, which more effectively inhibits parasite expression, and results in failure of the parasite immune suppression and a failed infection (lower right). Across these four host-parasite genotype combinations, the expression level of host and parasite genes are negatively correlated. This correlation is a result of indirect genetic effects: host expression down a column differs because of genetic variation in the parasite (host genotype being held constant); parasite expression varies along a row as a function of host genotype (parasite genotype being held constant). Figure created with BioRender.com.

Correlated gene expression between hosts and symbionts can also arise via within-generation selective filters (Combes 1991, 2001). Variation in defense or virulence (compatibility filters) or in things like behavior or diet (encounter filters), will determine which combinations of hosts and parasites end up living together (Anacleto et al. 2019; Kołodziej-Sobocińska 2019; Bolnick et al. 2020). Evolutionary biologists sometimes use simplified mathematical models to study coevolution that assume a "matching allele" process in which infection succeeds when a pathogen carries an allele that allows it to bind to host cell surface receptors and fails otherwise. An alternative 'gene-for-gene' model describes situations where a given host genotype (e.g., a particular MHC allele) can recognize and elicit a response to a particular pathogen genotype (thus blocking infection) (Bergelson et al. 2001; Woolhouse et al. 2002; Živković et al. 2019; Ebert and Fields 2020). For example, infection success of malaria in mosquitos is determined by the specific host genotype by parasite genotype interaction (Lambrechts et al. 2005), similar patterns have been found when comparing resistance in different genotypes of *Daphnia* to different bacterial isolates (Carius et al. 2001). Similar filter models exist for the matching of symbiotic mutualists, where certain host genotypes will recognize and support specific partner genotypes (Parker 1999; Stoy et al. 2020). Consider a population of hosts and symbionts, each of which is polymorphic for genes mediating this matching process (Fig 3). Even if host-symbiont encounters are random with respect to their genotypes (Fig 3A), only certain combinations of host-symbiont pairs will succeed in establishing and will be observed (Fig 3B). If genetic variation within each species imparts constitutive differences in gene expression, then symbiont gene expression will be correlated with the gene expression of its host and CSCoE will be observed (Fig 3C). Unlike the IGEs described above, this does not require plasticity of expression, nor any response of either species to the others' expression or genotype. Instead, these filters represent a direct case of selection in which the fitness of the symbiont or host depends directly on the interaction between their genotypes. Similar filtering can arise from genetic covariance in encounter rates, for instance, if host and parasite genotypes are both distributed non-randomly across microhabitats on a landscape resulting in non-random combinations of genotypes coming into contact. These filters can be difficult to detect without carefully controlled experiments, particularly because missing combinations will be difficult to quantify in nature, but they can still play a role in shaping gene expression correlations between hosts and parasites.

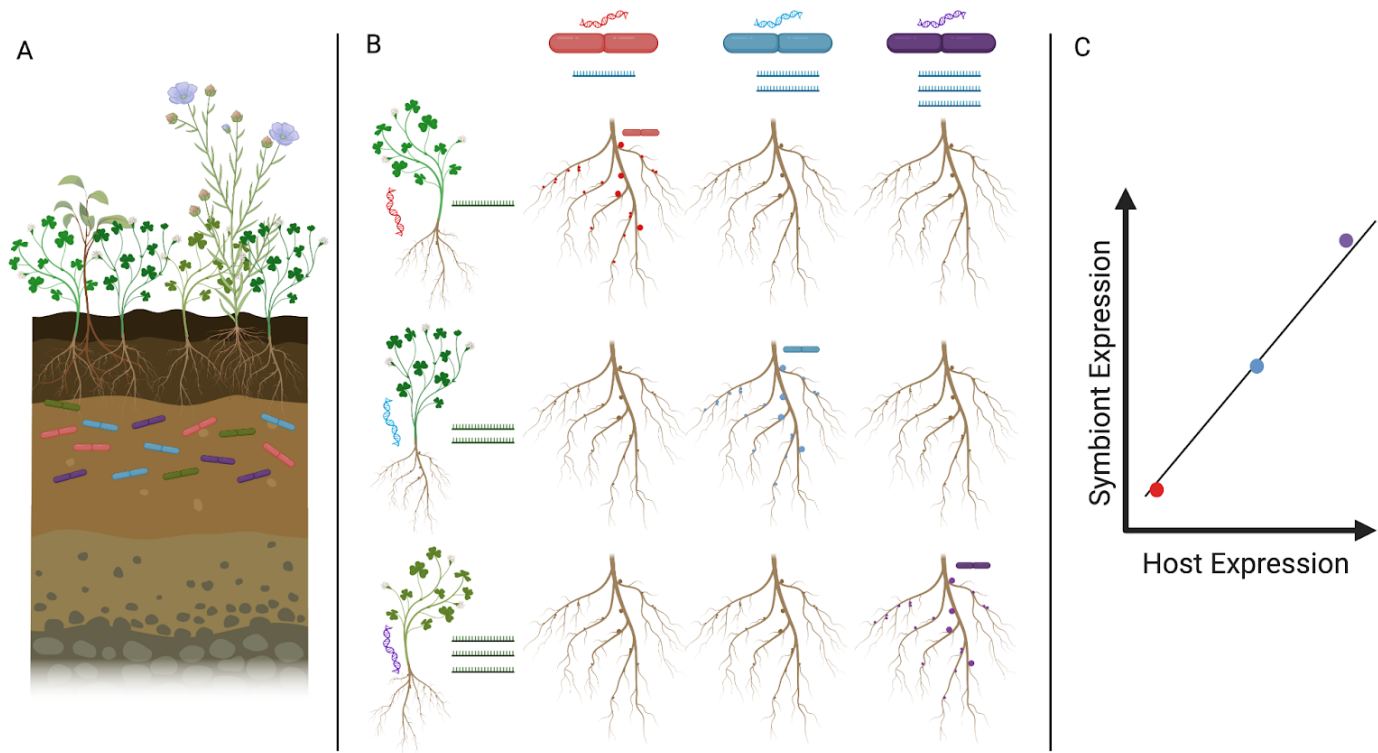

**Figure 3.** An illustration of a within-generation selective filter leading to correlated gene expression between hosts and symbionts. This example is portraying a mutualist microbe forming a symbiosis in the root nodules of a plant host. A population of several host genotypes, here represented by plants with different colors and shapes, is polymorphic for a matching allele, represented by different colored DNA strands. Different symbiont genotypes, represented here by different colored microbes, are also polymorphic for a matching allele. Even if host-symbiont encounters are random (panel A), only certain combinations of host-symbiont pairs will lead to successful establishment and be observed when sampling (panel B, colored nodules). If underlying genetic variation in hosts and symbionts also leads to variation in gene expression (represented by different numbers of mRNA for each genotype), correlations will arise and cross-species coexpression will be observed across the different matched host-symbiont pairs (panel C). In this case, fitness for both the host and symbiont depends on the direct interaction of their different genotypes. Figure created with BioRender.com.

Regardless of whether CSCoE is due to direct or indirect genetic effects, selection, or plasticity, coexpression analyses hold promise for identifying genes and pathways that affect the physiology and fitness of interacting species. The genes exhibiting coexpression across species are strong candidates as being, or being regulated by, major targets of coevolutionary selection. However, identifying causal genes under selection from gene expression data alone is challenging (Schadt et al. 2005; Kim et al. 2011; Schaefer et al. 2018). It is possible that the causal polymorphism driving expression covariance between species lies in the coding sequence of one gene (e.g., MHCIIb, which contributes to parasite detection), but actuates changes in the expression of genes further down the regulatory circuit (Fig 4). Thus, finding genes with correlated expression across species does not, in itself, provide direct insight into evolutionary process.

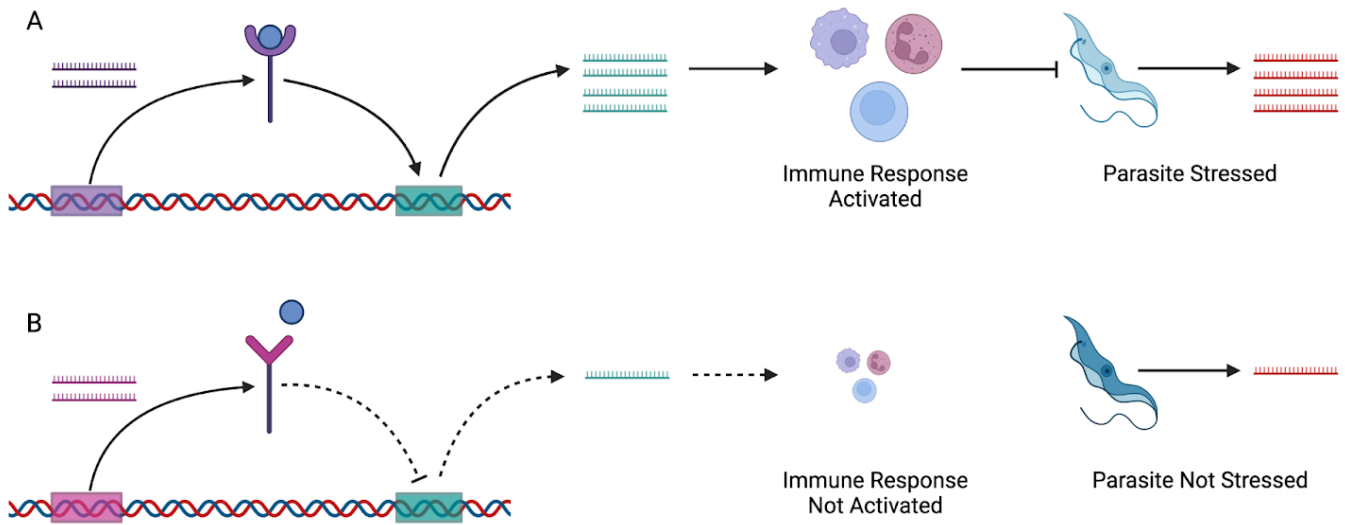

**Figure 4.** A diagram illustrating how a polymorphism in a causal gene can cause downstream changes in gene expression and correlations between host and parasite gene expression. In this example, sequence differences in a gene result in changes to a receptor used for parasite recognition. In row A, the receptor can bind to the parasite antigen, while in row B, the parasite remains undetected. The sequence difference in this host receptor gene then lead to changes in the expression of other genes further down the regulatory cascade, in this case, immune activation in row A but not in row B. Immune activation in row A then elicits gene expression changes in the parasite, which is now stressed. In this example, cross-species gene expression data and coexpression analysis would identify the second green gene as being potentially important for host-parasite interactions, but the causal gene under selection and likely contributing to coevolution (the first pink/purple gene) would not be detected with gene expression data alone. Figure created with BioRender.com.

**Biological and Temporal Scale**

CSCoE analysis can be carried out at any of several biological scales. At one extreme, coexpression at the level of individual host cells carrying intracellular parasites (Reid et al. 2018;

Hayward et al. 2021; Mukherjee et al. 2021) will focus on mechanistic details of parasite development and the molecular interactions driving parasite exploitation of the host cell, but has little to convey about evolution. Alternatively, CSCoE may be examined among genetically disparate hosts infected by different parasite genotypes (e.g., Fig. 1, Fig 2). The focus here is on how specific host-parasite pairs interact. Still larger scales of analysis may be used. For example, among-population covariation is a likely outcome of a geographic mosaic of coevolution (Thompson 2005) and of local adaptation; mechanisms of coevolution at this scale can be revealed using common-garden experiments with sympatric and allopatric host-parasite combinations (Feis et al. 2018). One could sample numerous host populations and characterize the average expression level of each host population, and the average expression level of each parasite population, then test for among-population CSCoE. At still larger scales, one expects gene expression to diverge between closely related host species, and among the co-speciating parasites that infect them (Hafner et al. 1994; Paterson and Banks 2001; Clayton et al. 2003). With a sufficiently large clade of both host and parasite species, one can generate phylogenetically corrected covariances between host and parasite species. We discuss the evolutionary insights from these different scales in the next section.

Across these biological scales, we must also consider the timing of sampling. Many comparative studies are limited to a single time point, yet host-parasite interactions are often characterized by a sequential series of events or phases (Torres et al. 2016; Hall et al. 2017b, 2019). Infections progress in a "step-wise" fashion from exposure to establishment to recovery/death and hosts move from parasite recognition to a response that often progresses through time (as with adaptive immunity). Sampling at different times during this process will capture different phases of the interaction (Foth et al. 2011; Tierney et al. 2012; Choi et al. 2014;

Colgan et al. 2020). Selection often targets key points in this progression, and our ability to differentiate these stages and correctly time our sampling will shape our understanding of the mechanisms underlying host-parasite interactions. Sampling through time or at different key steps in a comparative context will allow for a more thorough examination of which step populations differ and coevolution is occurring (e.g. (Hund et al. 2022)). Each of these steps can also represent a filter, where different host and parasite genotype combinations result in different outcomes (Fig 3).

**Coexpression analyses may identify candidate genes underlying variation in host-parasite interactions.**

In single-species coexpression analyses, researchers often look for sets of coexpressed genes (modules) because these genes are likely to be functionally related (e.g., in the same pathway) (van Dam et al. 2018). Extending this logic to cross-species coexpression, it is reasonable to infer a functional relationship between coexpressed host and parasite genes. In particular, we may be interested in identifying host and parasite genes underlying antagonistic coevolution (e.g., matching allele or gene-for-gene models of coevolution) (Märkle et al. 2021a). For coevolution to occur, both species must have genetic variation that impacts the outcome of infection and both species' fitness. It is likely that this genetic variation will lead to variation in gene expression, and perhaps coexpression between species. The question is, can we work backwards from the coexpression patterns that we observe to infer the genes driving coevolution?

It is not safe to simply assume that coexpressed genes are the target of selection and coevolution. If the coexpression network is constructed from data collected from wild-caught individuals, the variation captured in the coexpression network might reflect both species' response to a shared environment (e.g., temperature), rather than genetic variation. A second concern is that variation in gene expression might reflect polymorphism at a different gene (e.g., a trans-eQTL), which is itself not differentially expressed (Fig 4). Thus, cross-species coexpression might be more effective at identifying regulatory pathways engaged in coevolution, rather than finding causal polymorphic genes.

One solution would be to incorporate population genomic analyses looking for signatures of past natural selection on coexpressed genes (or, known upstream regulators of coexpressed genes). Indeed, combining gene coexpression data with genomic data has proven to be a powerful approach within species for understanding the functional relationship between genes and their evolution (Ruprecht et al. 2017; Mack et al. 2018; Randhawa and Pathania 2020; Yao et al. 2020). As reviewed by Ebert and Fields (2021), once genes of interest are identified, the nature of how coevolutionary selection has shaped their evolution can be characterized using population-genomic analyses. Establishing that candidate genes are actually involved in coevolution, as opposed to adaptation in just one of the two interacting species, would require evidence of selection in both species. Consequently, researchers may wish to test whether host-parasite gene coexpression is especially likely to involve genes, in both species, with strong signatures of selection. It could be equally interesting (indeed, surprising) to find that selection on cross-species coexpressed genes is no different than genome wide averages.

Evidence for selection is even stronger when the same gene is under selection in multiple replicate populations (parallel evolution). However, complex genotype-phenotype mapping can lead to non-parallel evolution (Bolnick et al. 2018) that makes adaptation harder to detect. Taking a network approach can help solve this problem: although selection might not act on the same gene in each replicate population, it may act on a set of closely interacting genes. Thus, one can test whether selection repeatedly acts on genes within a coexpressed module, even if it does not repeatedly act on any specific gene. In particular, we might ask whether selection is generally stronger on genes in cross-species modules, than in genes from species-specific modules. This logic also applies to polygenic traits. Selection tends to act weakly on individual genes contributing to quantitative traits. Genomic tests of selection often have weak power to detect selection on polygenic trait loci but taking a network approach might prove helpful. A coexpression module underlying polygenic trait adaptation may show an average signal of selection, or a concentration of genes with elevated signals of selection, even if no single gene stands out (Hämälä et al. 2020b).

      Although genomic analyses provide insight into evolutionary history, they may not reflect selection happening in contemporary populations, which require direct measures of fitness. To detect contemporary selection, researchers can use standard general linear model tools to test whether individuals' fitness depends on their own gene expression, or their antagonist's gene expression. For hosts, fitness can entail a complex mix of survival, lost resources, costs of immune responses, changes in foraging ability, predator evasion, mate attraction, and ability to generate and rear offspring. For a parasite, fitness can simply reflect success of infection (Fig. 3), or growth rate within a host and subsequent transmission to new hosts. As with any analysis of fitness, measures will often be limited to particular fitness correlates or components. The gene-gene correlation matrix (Fig. 1C) can be decomposed into major axes of variation (e.g., eigenvectors from Principal Component Analysis), and host-parasite pairs can be quantitatively arrayed along each axis. Then, fitness can be regressed on the coexpression eigenvalue scores. A crucial challenge, however, is that parasite fitness may entail failed infections, and one cannot

measure gene expression (or host-parasite correlations) on an absent parasite (see Methodological Considerations, below).

**Cross-species coexpression may reveal variance in ecological interaction strengths.**

The field of ecology has long emphasized species-level analyses: tracking changes in species densities, food webs showing which species consume others, or analyses of species-level diversity. Yet, in the past decades, there has been growing appreciation for the often substantial ecological diversity within species, or even within populations (Forsman and Wennersten 2016; Feiner et al. 2019; Auer et al. 2020; Chardon et al. 2020; Shaw 2020). There may be among-individual variation in diet preferences, parasite susceptibility, predator evasion, metabolic rate, stoichiometry, movement and more. Cross-species coexpression networks are only detectable when there is among-individual trait variation among individual hosts, and among individual parasites. Thus, finding cross-species expression directly implies that there is likely to be intraspecific variation in ecological interactions For instance, host plants and their symbiotic mycorrhizae can exhibit among-individual variation in the benefits each provides to the other (Deveautour et al. 2021).

Individual variation in parasite susceptibility and exposure, or the ability to support or suppress the growth of different mutualists, will shape the assembly of symbiont communities within each host. This idea has been well-demonstrated with research into microbiome communities (Hall et al. 2017a; Datta et al. 2018; Luca et al. 2018; Scepanovic et al. 2019; Bubier et al. 2021). While the study of coevolution and the use of cross-species gene coexpression has primarily focused on single pairwise interactions between hosts and symbionts, in nature, hosts typically harbor a diverse and dynamic symbiont community. This can lead to complex and higher-order interactions between the host and different symbionts and between the symbionts themselves, often mediated through host physiology or immunity (Rigaud et al. 2010; Bordes and Morand 2011; Buhnerkempe et al. 2015). For example, a parasite that suppresses host immunity can facilitate the growth of other parasites, while a parasite that tolerates the immune system (and thus activates it) can increase immune costs for other infections. However, whether co-infecting species are affecting the same or different aspects of host physiology, or imposing tradeoffs, is often unknown.

Expanding CSCoE studies to include multiple symbionts offers a new and exciting approach to begin teasing apart these multi-species interactions (e.g., a multi-species version of Fig. 1). This approach could allow us to monitor or uncover mechanisms for how coinfecting species are facilitating or competing with one another within hosts and how the symbiont community is collectively shaping host state. It may also allow us to measure the extent to which coevolution is pairwise or diffuse (Hall et al. 2020; Agrawal and Zhang 2021). If many symbiont species are represented in a host-parasite coexpression module, then there is greater opportunity for pleiotropic effects in which one parasite modifies host physiology in ways that indirectly impact other species – resulting in diffuse coevolution. By contrast, if there are multiple modules containing both host and parasite genes, but each of these modules contains genes from only one parasite lineage, then coevolution would be pairwise. Multilayer network analysis is particularly well suited to accommodate this type of data, in which the gene coexpression network for each of multiple species can be represented as its own network layer with inter-layer connections representing cross-species expression correlations (De Domenico et al. 2015; McGee et al. 2019; Škrlj et al. 2019; Bazzi et al. 2020).

**Methodological considerations and future directions**

There are a variety of experimental opportunities afforded by the study of CSCoE. The most basic approach (already articulated) is to experimentally stage factorial combinations of varying host genotypes interacting with varying parasite genotypes (e.g., Fig. 1A, Fig. 2, Fig. 3). One can then test for direct and indirect genetic effects by differential expression analysis of transcriptome data (of each species) as a function of host genotype, parasite genotype, and their interaction. By measuring the frequency of successful host-parasite combinations in these experiments, one can distinguish between IGE and filtering mechanisms. If coexpression arises from IGEs, each species' expression will vary as a function of the genotype of its partner species (Fig. 2). If coexpression arises entirely from selective filtering (Fig. 3), a non-random subset of host-parasite genotype combinations will be observed, but parasite expression might be exclusively a function of parasite genotype, and host gene expression exclusively a function of host genotype. Thus, observational coexpression results cannot be used to infer a causal mechanism; experimental approaches are needed for this.

Many experiments that look at host and parasite gene expression will also include data from uninfected hosts, or of symbionts in culture (Mohamed et al. 2020). Comparisons between infected and uninfected hosts (or between symbionts in the host and those in culture), are commonly used to understand plastic changes in gene expression or network structure that are associated with infection (Amar et al. 2013). One might assume that these differentially expressed genes are also likely to be the genes underlying host-parasite interactions, but this does not have to be the case. With studies that involve multiple genotypes of host and parasite, the genes that explain variation among infected individuals, or are associated with differential connectivity within the infection treatment, may not be the same genes that are differentially expressed between infected and uninfected hosts. For example, Mateus et al. 2018 found that while highly conserved genes were largely responsible for symbiosis establishment in cassava, most of the plant and fungal transcriptional responses after establishment were not conserved and were greatly influenced by the specific combinations of plant and fungal genotypes. This is not to say that data from uninfected hosts is not useful. In fact, it is interesting to directly test if genes that are differentially expressed between infected and uninfected hosts are also genes that are highly connected in coexpression networks and have expression that covaries with parasite genes. Data from uninfected hosts can also serve as an initial time point when studying how host-parasite interactions change during infection, and for understanding baseline differences in hosts when comparing host-parasite interactions among populations (Lohman et al. 2017).

      While cross-species gene expression data can capture expression dynamics that are coordinated between hosts and parasites, it remains difficult to determine the directionality (cause-and-effect) of these relationships. Genetic mapping provides a powerful approach to gain insight into that directionality, as it can connect allelic variation to phenotypic variation (gene expression levels) both within and across partner species. In other words, it allows researchers to identify polymorphisms in one species that influence the gene expression of the partner species. One approach is to generate genetically distinct lines or crosses of one species (for mapping) while keeping the genotype of the partner species constant in experimental lab infections (Saeij et al. 2007; Wu et al. 2015; Guo et al. 2017; Soltis et al. 2020). Given consistent environmental conditions and timing, this design allows for the ability to map the gene expression phenotype of both species to genetic markers in the variable species, using standard eQTL methods (Gilad et al. 2008; Nica and Dermitzakis 2013). This approach has now been used in several model host-parasite systems and results have provided clear evidence that the genotype of one species can

impact the gene expression of its partner, and have identified candidate genes for interspecific signaling (Saeij et al. 2007; Wu et al. 2015; Guo et al. 2017; Soltis et al. 2020). However, many of these studies use host or parasite populations with limited genetic variation, as is often true in lab model systems.

A more ambitious experimental design could include factorial combination of varying host and parasite genotypes, allowing mapping to occur simultaneously in both species. This could be done by taking immunologically divergent populations of hosts and interbreeding them to generate recombinant F2 generation hybrids, then doing the same for divergent populations of a sexually reproducing parasite. Researchers could then stage many experimental infections of F2 hybrid hosts exposed to F2 hybrid parasites, and measure gene expression on the resulting infected hosts and their successful parasites. By genotyping all individuals for numerous genetic markers, one could then build both host and parasite linkage maps in which to carry out within and cross-species eQTL mapping. With a sufficiently large sample size, one could even map epistatic interactions among host genes, among parasite genes, or between host and parasite loci. This endeavor requires genetic variation in each species, with species amenable to genetic crosses and experimental infections. While it represents an imposing experimental task, it offers the opportunity to identify the regulatory genetic basis of between-species interactions.

Reciprocal cross-species eQTL mapping experiments are rarely tractable and are likely limited to a few research systems. The alternative is to use naturally occurring combinations in polymorphic wild populations and perform cross-species genome-wide association studies (co-GWAS) (Nuismer et al. 2017; Ebert 2018; MacPherson et al. 2018; Wang et al. 2018). This method searches for cross-species allelic associations between parasite and host SNPs and has typically been used to detect significant correlations with infection outcomes. If there is covariation in gene expression between interacting species, one can sample a large number of naturally occurring host-parasite pairs and obtain gene expression and whole genome sequencing data for both interacting species. Using co-GWAS approaches (Nuismer et al. 2017; Ebert 2018; MacPherson et al. 2018; Wang et al. 2018; Märkle et al. 2021b), one can then test for relationship between either (or both) species' gene expression, and genotype at both host and parasite genetic markers. Cross-species gene regulation would be revealed by instances where, for example, host gene expression was a function of parasite genotype while controlling for the effects of host genotype. The standard statistical problem of multiple comparisons is amplified in co-GWAS when association values are calculated for all possible comparisons of host and parasite SNP pairs (Märkle et al. 2021b). This issue can be mitigated somewhat, because rather than testing for all expressed transcripts' dependence on all genetic markers in two species, one can first narrow the focus to transcripts that show CSCoE. Coexpression analyses can also draw attention to modules which can be analyzed as a unit rather than as individual genes, again reducing the multiple comparison problem.

Another interesting possibility would be to conduct multi-scale analyses of coexpression, using both among-individual and among-population sampling strategies. With interacting species, among-individual variation is most likely to reflect either IGEs or selective filtering (as noted above), whereas among-population covariance adds in the potentially substantial effect of evolutionary divergence in one or both species. If the individual scale indeed reveals interacting host and parasite genes that are likely to be under strong coevolutionary pressures, then we would expect that these same genes contribute to among-population covariance as well. To exclude the confounding effect of environment in this covariation, it is advisable to examine among-population gene coexpression in hosts and parasites in a controlled 'common-garden' environment. For example, (Will et al. 2020) quantified gene expression in laboratory infections of related species pairs of ants and cordyceps fungi. Taking this comparative approach allowed the authors to identify differentially expressed genes underlying potential host behavioral manipulation strategies that are shared across different cordyceps – ant interactions.

To confidently attribute observed variation to a candidate gene, the gold standard for proof is to do gene editing (Bak et al. 2018; Anzalone et al. 2020; Broeders et al. 2020). Experimentally altering a focal DNA sequence, then observing the expected phenotypic change, provides clear evidence that the altered DNA has a direct causal role in generating that phenotype. The same principle applies to cross-species coexpression: ultimately, we would like to see experimental genetic evidence that a particular DNA sequence in one species has a causal effect on gene expression in its partner species (Tierney et al. 2012; Schulze et al. 2016). Gene knockouts of one partner or the other in laboratory model systems have done this successfully to some extent (Tierney et al. 2012), and tools like CRISPR/cas9 gene editing are democratizing functional genetics in a wider variety of organisms. It is now technically feasible to edit a candidate gene in a group of animals (e.g., hosts) and test the effect of this alteration on expression in their parasites, or on host-parasite coexpression.

**Challenges associated with cross-species expression experiments and analysis**

While sequencing costs have decreased, large experiments that include gene expression from both hosts and parasites still represent a substantial expense. Researchers must make decisions about how many samples can be included, which imposes choices and tradeoffs that require care in the design of studies, and caution in the inference we can draw from them. For example, increasing variation for hosts and parasites beyond a single genotype inherently increases sample size. One approach is to vary the genotypes of just one of the partner species while using a single genotype for the other (Saeij et al. 2007; Wu et al. 2015; Guo et al. 2017; Soltis et al. 2020). However, this limits our ability to study coevolution, which relies on epistatic interactions between varying host and symbiont genotypes, as we would find in natural populations.

Sampling costs (and other logistics), also limit the number of time points that can be included in experiments, yet we know that timing in these interactions can be very important. Sampling at only one time point may miss key steps that are serving as filters (Fig. 3), at different developmental trajectories, or along the time course of an infection (Hall et al. 2017b). Good background knowledge of the system and its natural history will be essential to choosing timepoints, or stages of the interaction, that are likely to be informative.

Host-parasite interactions are also shaped by the ecological context in which they occur (Wolinska and King 2009; Morand and Krasnov 2010; Duffy et al. 2012; Gervasi et al. 2015). Things like resource availability, coinfection, and other stressors will impose tradeoffs that change host investment in parasite defense. Ecology will influence which hosts are exposed to which parasites, and host density will influence transmission dynamics. On larger scales, environmental factors often shape the distribution of genotypes on the landscape and will impact which host-symbiont genotypes interact (Morand and Krasnov 2010; Fountain-Jones et al. 2018; Becker et al. 2020). Experiments that can study cross-species gene expression in different environmental contexts will be key to understanding how ecology shapes host-parasite interactions and coevolution.

Once the data are in hand, the analyses associated with CSCoE are almost entirely correlational. Because of the magnitude of data being analyzed, it is inevitable that correlations will be found, and the cutoffs, assumptions, and decisions made during analysis will all influence the results, as is the case in the analysis of single-species gene expression data (Lovén et al. 2012; Costa-Silva et al. 2017; Chowdhury et al. 2020). For example, many programs used to build gene coexpression networks have been designed for building networks within a single species, and it is currently an open question whether we should use different assumptions and defaults when applying these same tools to multiple species. At the end of the day, what we call network structure is an artificial construct. How realistic these models are and their biological relevance will depend on experimental design and careful analysis and interpretation. As with any study, applying findings beyond the specific genotypes and environmental conditions used for the experiment needs to be done with caution.

**Conclusion**

Incorporating cross-species gene expression and network analysis to the study of host-parasite interactions has the potential to improve our understanding of the ecological and evolutionary processes shaping these interactions. The increasing accessibility of sequencing approaches and new analytical tools will expand our ability to study enduring and fundamental questions about host-parasite interactions and to understand the complexity of these interactions in new ways. These approaches open up new experimental design possibilities including: combining gene expression data and genomic data across interacting species, functional tests of gene targets or pathways uncovered in cross-species networks, how network structure is rewired over the course of an infection, and using gene expression data to study interactions between multiple symbionts within the same host. Many of the ideas and approaches discussed in this paper could also be applied beyond host-parasite interactions to study other symbiotic relations, and even to study other situations where individuals interact closely for extended periods of time, such as within social groups (Vojvodic et al. 2015; Bailey et al. 2018) or between parents and offspring (Groothuis et al. 2019; Xavier et al. 2019). While coexpression network analyses are largely correlational and experiments must be designed with care, they offer a powerful new framework and toolkit for advancing our understanding of the evolution and ecology of host-parasite interactions.

**<u>Box1 – Gene Coexpression Networks</u>**

Coexpression networks provide an efficient way to summarize information on how the expression of genes covaries among samples. The nodes of a network are genes and a connection between two nodes (genes) denotes a joint expression of the two genes. The samples from which expression data are collected (for within species networks) can be different tissues within an individual, stages of development, individuals living in different environments, or genotypes living in the same environment. Given that genetic variation is central to evolution, coexpression networks constructed from expression data of multiple genotypes within a population or species are likely most relevant for understanding coevolution.

The standard approach for constructing a coexpression network is to calculate the correlation coefficient between the among-sample expression profiles of each pair of genes, i.e., $Cor(x_i, x_j)$ where $x_i$ and $x_j$ are vectors of expression levels of each of N expressed genes across each of M samples). This results in an NN adjacency matrix, with each element being the pairwise correlation coefficient of the corresponding genes. For correlation to be high, the expression of two genes must covary across the M samples. Importantly, if two genes both have high expression in all M samples, or if one of the two genes has high expression while the other's is low in all M samples, then the correlations between those genes will be near zero and their connectivity will be low. Not surprisingly, in practice there are many decisions that need to be made in constructing this matrix including standardization of data, choosing the correlation measure, opting for signed or unsigned networks, and choosing a filtration method if correlations truly must be expressed in a binary way, as significantly coexpressed or not.

Coexpression networks are related to, but distinct, from differential expression analyses. Differential expression analyses compare the expression of genes between two samples to identify genes that are upregulated or downregulated, in response to an environmental stimulus, when comparing different tissues, or when comparing different species, and so on. Importantly, DE analyses are focused on identifying differences in the mean expression of genes instead of their interactions. From the perspective of host-symbiont relationships, DE analyses might be useful to understand how an organism responds to infection. These analyses are often conducted by comparing the expression of each gene pairwise, but multivariate approaches (e.g. PCA) could be applied to reduce the dimensionality of the expression data before comparing between sampling groups.

—-------------------------------------------------------------------------------------------------

**Box 2  Cross-Species Coexpression Networks in Practice**

Coexpression networks are routinely constructed for single species, using statistical tools such as WGCNA (Langfelder and Horvath 2008) to compute gene coexpressions and analyze the resulting networks (for instance to identify coexpressed modules). Expanding these approaches to examine data from two species, i.e., a host and a parasite, can largely be done using existing tools. For example, one might simply feed the gene expression data through the same pipeline, changing from N genes in a single species, to NH host genes and NP parasite genes, and then characterizing edges as being host-host, parasite-parasite, or host-parasite (i.e. interlayer).

Some additional technical choices must be made and careful attention to statistical details is necessary for the construction of a CSCoE network. For example, since average gene expression levels are likely to be much lower in parasites compared to hosts, when using dualRNAseq approaches, one may need to normalize counts for hosts and parasites separately before computing correlations, or possibly use a different measure of association between genes that is not sensitive to scale, such as mutual information (Cover and Thomas, 2006). It is then advisable to work with the correlation data directly if possible. From there, dichotomizing the association data to obtain binary relationships (on or off, coexpressed or not), introduces two problems that are particularly pronounced for CSCoE network analyses. First, one should approach data with the expectation of finding very few cross-species dependencies, which means choosing significance level that ensure an unusually low false discovery rate. This could lead to within species networks that are sparser than they need to be. Second, dichotomizing correlations can introduce spurious structure in the network (Cantwell et al., 2020). With two species, these effects may manifest themselves on different scales for each species and for the interaction between them, leading to more biases. Sample sizes should also be chosen carefully to ensure sufficient statistical power to detect cross-species coexpression, considering the variance among replicates and the researchers' desired ability to detect a particular effect size.

The construction process for CSCoE networks is similar to the standard procedures in gene coexpressions network analysis—with the additional consideration highlighted above. However, because these networks are inherently multi-layered (with interaction data from each species corresponding to a layer and inter-layer and connections representing correlations between the expression of genes from different species), multi-layer tools (Kivelä et al. 2014) need to be used to for module identification and annotation (Huang et al. 2021), or the ranking of gene importance or centrality (Bianconi 2021).

For module identification, one has to define a test statistic denoting module quality. Of particular interest in coexpression networks are dense modules with many internal connections, indicating groups of covarying genes. Such modules could be found with descriptive methods such as multi-layer analogs of modularity maximization or, even better, with inferential methods that assign a likelihood to separations in modules (Zhang 2020, Peixoto, 2021). Generative models with covariates can be useful for this purpose (Razaee et al. 2019; Contisciani et al. 2020). As mentioned in the main text, the content of modules will be indicative of interactions between species. For example, if we look for modules of densely connected genes, modules with only hosts or only parasite show covariation only in either species, while mixed modules will reveal rich interactions between the two species.

—-------------------------------------------------------------------------------------------------

**Author Contributions**
AKH, PT, and DIB conceptualized the initial framework of the manuscript. JGY contributed expertise on network statistics and analysis. AKH and DIB made the figures. All authors contributed to discussing ideas, writing, and editing the manuscript.

**Acknowledgements**
Thank you to members of the Bolnick and Snell-Rood lab groups and to R. Runghen for their helpful feedback on drafts of this manuscript. AKH was supported by a James S. McDonnell postdoctoral fellowship and funding from the Templeton Foundation (award 62220 to E. Snell-Rood at the University of Minnesota). PT was supported by funding from the National Science Foundation (NSF IOS-1856744).

**Data Accessibility Statement**
There is no data associated with this perspective.